# Software paper for submission to the Journal of Open Research Software

## (1) Overview

### Title
PyRDM: A Python-based library for automating the management and online publication of scientific software and data

### Paper Authors
1. Jacobs, Christian T. (*corresponding author*)
2. Avdis, Alexandros
3. Gorman, Gerard J.
4. Piggott, Matthew D.

### Paper Author Roles and Affiliations
1. Department of Earth Science and Engineering, Imperial College London, London SW7 2AZ, UK

2. Department of Earth Science and Engineering, Imperial College London, London SW7 2AZ, UK

3. Department of Earth Science and Engineering, Imperial College London, London SW7 2AZ, UK

4. Department of Earth Science and Engineering, Imperial College London, London SW7 2AZ, UK

### Abstract
The recomputability and reproducibility of results from scientific software requires access to both the source code and all associated input and output data. However, the full collection of these resources often does not accompany the key findings published in journal articles, thereby making it difficult or impossible for the wider scientific community to verify the correctness of a result or to build further research on it. This paper presents a new Python-based library, PyRDM, whose functionality aims to automate the process of sharing the software and data via online, citable repositories such as Figshare. The library is integrated into the workflow of an open-source computational fluid dynamics package, Fluidity, to demonstrate an example of its usage.

**Keywords**
scientific software, data, automated publication, reproducibility, recomputability, Python, Figshare, GitHub

**Introduction**
The 21st Century has seen a massive growth in the quantity and complexity of data produced by scientific software. At present, such data is typically confined to the researcher's computer for analysis, and eventually condensed down into a set of key results and figures for publication in a journal. The raw data itself is rarely made available to the wider community. Furthermore, the software (which can itself be a research product) is often not shared because of the current lack of intrinsic motivators for the researcher who developed it, thereby preventing the data from being recomputed by others. In academia the impact factor of an individual is commonly measured by the number of publications and their associated citation counts; making the software available can be seen as an additional effort that will not reap significant rewards for the researcher since they will gain little recognition from something which is not viewed in the same regard as a journal article. Even in the case where the software *is* made available, it is often poorly referenced; usually either the user manual or the software's website are cited, but neither resource is satisfactory since they do not provide information about the specific revision of the source code that was used to produce the data. Without the means of replicating and verifying the correctness of a result, which requires full and open access to both the source code and raw data [1,2], its credibility could be considered questionable [3]. Reproducibility and recomputability are therefore becoming increasingly recognised by scientific communities as measures of research quality.

With the release of new tools such as Fidgit [4], the Mozilla Science Lab's "Code as a Research Object" digital object identifier (DOI) generator for GitHub repositories [5], and the dvn [6] and rfigshare [7] packages which interface with the Dataverse Network (www.thedata.org) and Figshare (www.figshare.com), publishing software and data online is becoming increasingly straight-forward. However, care must be taken to ensure that (a) the version of the software that is published is exactly the same as the version that generated the data, (b) that any provenance metadata (e.g. the DOIs of the software and input data) is appended to the output data before being published, and (c) that all necessary files are selected for publication (i.e. any collected/measured input data such as bathymetry for an ocean simulation, and any output data such as field solutions and plots). Doing so by hand can be a complicated, time-consuming and error-prone task.

In order to further encourage the publication of software and data online, we have developed an open-source Python-based library called PyRDM which provides the capability to integrate research data management into the workflows of scientific software packages, so that the software and data can be curated at the push of a button. The functionality of PyRDM is encapsulated in several Python modules which, in short, facilitate the automated publication of scientific software and data via online, citable repositories. Specifically, the Figshare online repository service is used to store the software and data, and provides DOIs for both resources so they can be appended to any provenance metadata. Not only does this allow a specific version of the software or data to be properly cited in a journal publication, it can also reduce data duplication by enabling the sharing of data with colleagues around the world. From the point-of-view of an individual researcher, being able to cite the code and data using a DOI brings reputation-related benefits since these outputs are often included in online publishing records (e.g. in Google Scholar profiles and *h*-indexes). Furthermore, the design philosophy of PyRDM is such that it is something that 'just works' from a user perspective, requiring minimal effort past the initial integration step.

One application of PyRDM has involved its integration into the workflow of Fluidity [8,9], an open-source computational fluid dynamics code. For a given simulation, users only need to enable a 'publish' option in the simulation's setup file (see Figure 1) and provide a minimal amount of information concerning which files they wish to publish (e.g. "*.vtu" will publish all VTK-based data files). All other issues such as determining which software version was used to run the simulation; checking whether that version has been published already (so its associated DOI can be reused); affiliating all authors of the software with the Figshare repository; appending

DOIs to the provenance metadata; and using MD5 checksums to determine which data files do not need to be re-uploaded to the Figshare servers each time a new version of the data is published (using the same DOI), are all handled automatically with the help of PyRDM. This is discussed in greater detail in the next section.

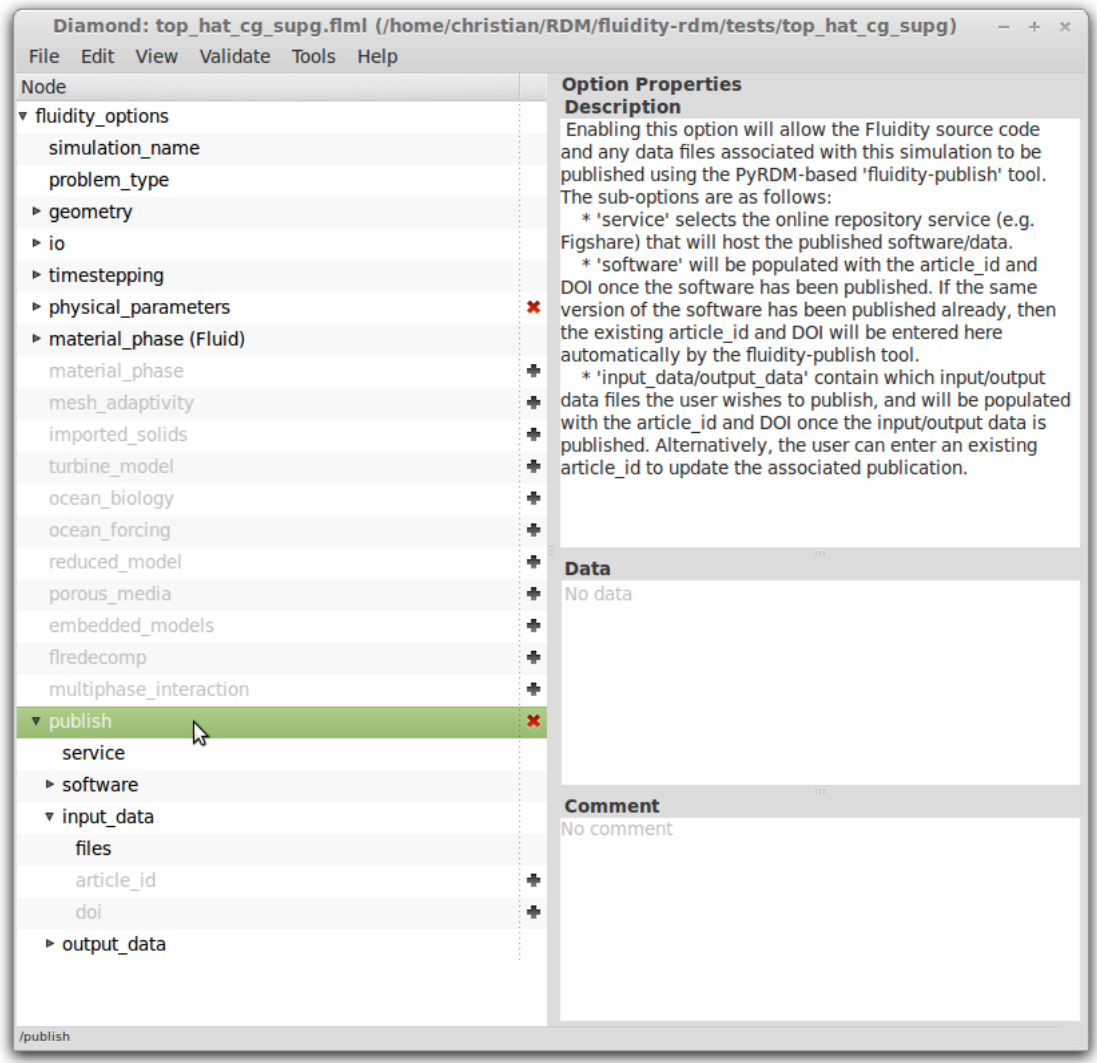

Figure 1: The 'publish' option enabled in a Fluidity simulation setup file. The file has been opened in the XML file editor, Diamond [10]. The description pane on the right-hand side describes the various sub-options available.

**Implementation and architecture**

PyRDM has been implemented using the Python programming language, and its design is centred around two classes, Publisher and Figshare. The Figshare class handles all the communication with the Figshare servers via the Figshare API to create code repositories and datasets, and upload any files. The Publisher class handles any tasks that are specific to the publication of software and data, such as determining which files are to be uploaded and automatically associating all authors with Figshare author IDs to the software repository. This is illustrated by the UML diagram in Figure 2. Note that the use of ORCID researcher IDs will likely be the preferred approach for affiliating authors in the future.

The publication of software (with a known version) is performed using the publish_software method of the Publisher class. The method first searches Figshare to determine whether the same version of the software has already been published. This version is based on a unique SHA-1 hash corresponding to a specific revision of the software in its repository; currently, only public Git repositories hosted by GitHub are supported. If the specific version of the software has already been published, then the method returns details of the existing publication (including the DOI). Otherwise, it creates a new code repository on Figshare, downloads the source code (with the specified SHA-1 hash) from the software's Git repository as a .zip file, and then uploads it to Figshare. The SHA-1 hash is then added as a 'tag' to the Figshare publication for future search and identification purposes.

To ensure that all authors are properly attributed to the software's publication, the publish_software method also looks for an AUTHORS file and, if present, parses it for Figshare author IDs and automatically adds any additional authors. Note that only the software's name, SHA-1 hash, location of the local Git repository containing the software, and the category of the software (e.g. "Computational Physics") are passed as arguments to publish_software; all other parameters, such as the Figshare authentication keys and location of the GitHub repository online, are obtained from the PyRDM configuration file which is populated just once by the user and by interrogating the local Git repository.

The publication of input and output data is performed using the publish_data method of the Publisher class. When the method is called, it creates a new fileset on Figshare and uploads the files whose paths are specified (along with a title and description of the data) in a dictionary of parameters which is passed in by the calling program (i.e. a software-specific publishing tool, explained below). It then returns details of the fileset, including its ID and DOI. Alternatively, PyRDM can be provided with a known fileset ID to update any existing data which creates a new version of that fileset on Figshare, but keeps the same DOI.

In the event of having to update a published fileset or create a new version altogether, it would be inconvenient to re-upload any files that have not been modified since the first time they were published. This is especially true if the files are large; examples include computational meshes, forcing data, and any field values used as initial conditions. Therefore, PyRDM only performs *selective* updates of files. The first time a file is uploaded to Figshare, a corresponding MD5 checksum file is created and stored locally. The next time the user attempts to upload that file, PyRDM determines whether the MD5 checksum of the file (in its current state) is the same as the MD5 checksum stored in the corresponding checksum file. If it is the same, the file is not re-uploaded.

The PyRDM modules can be imported and used in a separate Python program which acts as the automated publishing tool for a specific piece of scientific software. An example which considers the computational fluid dynamics code Fluidity can be found in the 'bin' directory of PyRDM. In this example, publication is viewed as a staged process; users may publish the software and input data first, run the simulation and then publish the output data. Alternatively, users can perform all of the publishing steps once the simulation has finished. When executing the Fluidity-specific publishing tool, the user only needs to provide the location of the XML-based configuration file of the simulation being considered (for details concerning the location of the input/output files) and what they would like published (software, input data or output data). At each stage, the Figshare code/fileset ID and DOI information are stored in the simulation's configuration file. Users can also provide PyRDM with an existing ID and/or DOI through the configuration file. When publishing output data, the DOIs of the software and input data used to produce that output data are appended to the simulation output's metadata to ensure data provenance. An example is given below:

```
<constant name="FluidityVersion" type="string" value="1baf80aac1e7e735b1cf182bc20761a0c6df7767"/>
<constant name="SoftwareDOI" type="string" value="http://dx.doi.org/10.6084/m9.figshare.1035081"/>
<constant name="InputDataDOI" type="string" value="http://dx.doi.org/10.6084/m9.figshare.1035083"/>
<constant name="CompileTime" type="string" value="May 23 2014 15:22:23"/>
<constant name="StartTime" type="string" value="20140523 154857.775+0100"/>
```

**Quality control**
Unit testing is carried out on the separate modules within PyRDM to verify their correctness. In particular, tests within the Figshare module create a temporary and private dummy dataset on the Figshare servers (without minting a new DOI), perform various operations on that dataset via the Figshare module's methods (e.g. adding a file, searching for that dataset, and adding a tag), then checks via assertion statements that the dataset has been modified in the correct way. The dummy dataset is then deleted after the unit tests have taken place. The tests within the Publisher module verify that the MD5 checksum functionality does indeed correctly determine which files need to be re-uploaded, and that the PyRDM AUTHORS file is correctly parsed for the Figshare author IDs.

Before publishing to Figshare, a trial run of the Fluidity-specific publishing tool which uses PyRDM was performed using data from a simulation of a 'top hat' advection problem (see [9] for more information) in order to test its usability and uncover any bugs that were not detected by the unit tests. The automated publication of the software, input data and output data was successfully achieved, yielding the following DOIs:

Software: http://dx.doi.org/10.6084/m9.figshare.1035081
Input data: http://dx.doi.org/10.6084/m9.figshare.1035083
Output data: http://dx.doi.org/10.6084/m9.figshare.1035093

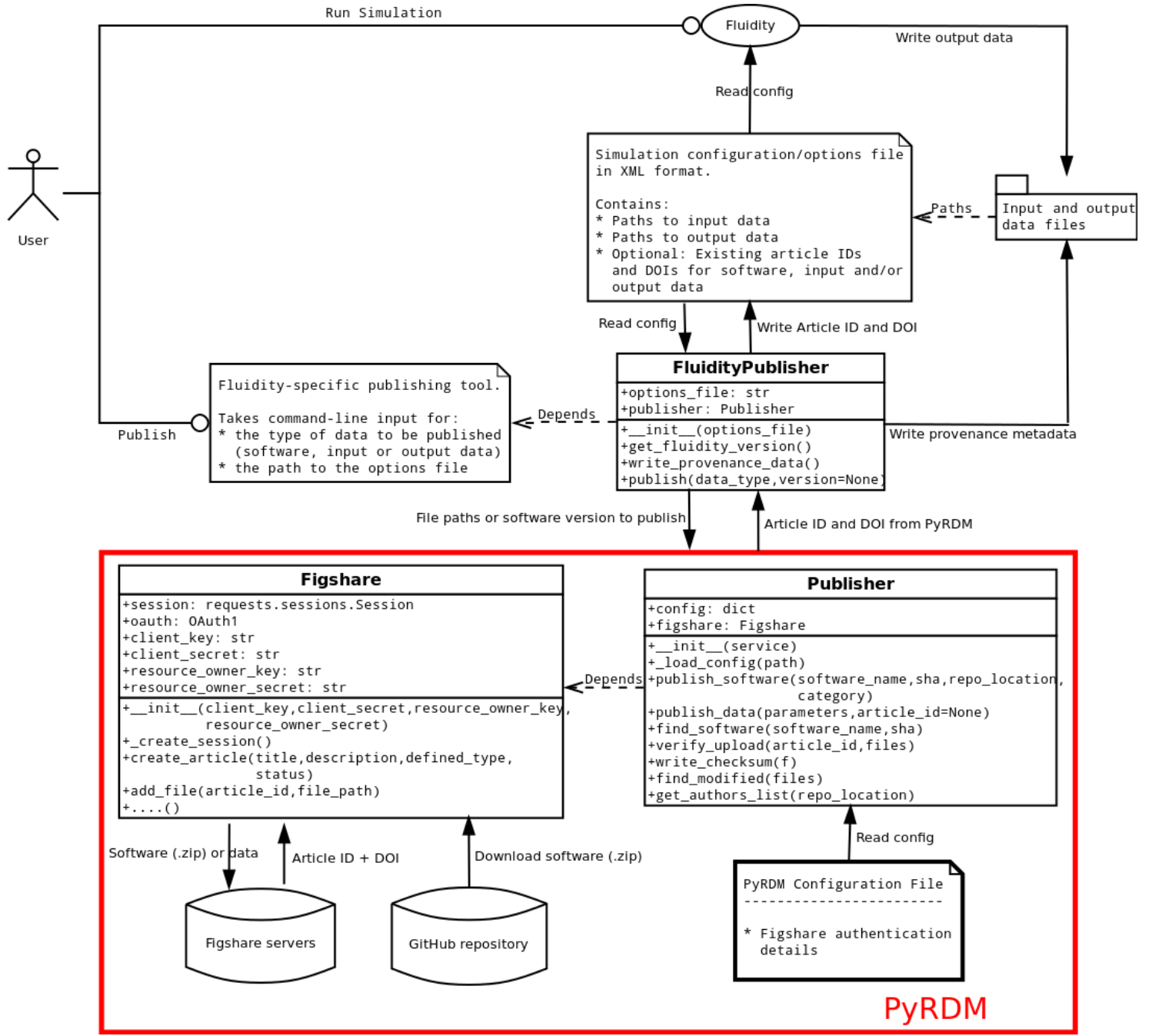

Figure 2: A UML diagram of PyRDM, illustrating how it fits into the workflow of the computational fluid dynamics code, Fluidity. Only the most relevant methods are listed in the classes. The term "article ID" refers to the ID of either a code or data fileset publication on Figshare.

## (2) Availability

**Operating system**
Linux (tested on Ubuntu Precise (12.04) 32-bit and 64-bit)

**Programming language**
Python 2.7.3

**Additional system requirements**
The computer will need enough free disk space to accommodate a complete download of the scientific software from GitHub, in order to then upload it to Figshare. A connection to the Internet will also be required.

**Dependencies**
- requests-oauthlib >= 0.4.0 (https://github.com/requests/requests-oauthlib)

- GitPython >= 0.3.2.RC1 (https://pypi.python.org/pypi/GitPython/)

- pdfLaTeX >= 1.40.10-2.2 (http://www.tug.org/applications/pdftex/), to build the user manual from the LaTeX source file.

**List of contributors**
1. Jacobs, Christian T. (Lead developer)
2. Avdis, Alexandros (PI/Supervisor, Developer)
3. Gorman, Gerard J. (PI/Supervisor)
4. Piggott, Matthew D. (Supervisor)

**Software location:**
   **Archive** (e.g. institutional repository, general repository) (required)
      **Name:** GitHub
      **Persistent identifier:**
      https://github.com/pyrdm/pyrdm/releases/tag/v0.1
      **Licence:** GNU General Public License (Version 3) (http://www.gnu.org/licenses)
      **Publisher:** Christian T. Jacobs
      **Date published:** 27/05/2014
   **Code repository**
      **Name:** GitHub
      **Identifier:** https://github.com/pyrdm/pyrdm

    ***Licence:*** GNU General Public License (Version 3) (http://www.gnu.org/licenses)
    ***Date published:*** 27/05/2014
  **Emulation environment**
    ***Name:*** N/A
    ***Identifier:*** N/A
    ***Licence:*** N/A
    ***Date published:*** N/A

***Language***
English

**(3) Reuse potential**
A long-term vision for PyRDM is for it to be integrated into a wider range of scientific workflows. Doing so will provide scientists with the means and motivation to publish their research outputs in a way that is beneficial to themselves (with respect to impact factor and citation counts) and the wider scientific community (with respect to recomputability and reproducibility). Much like the "Code as a Research Object" browser plugin [5], which places a 'publish' button on the GitHub webpage to allow developers to easily obtain a DOI for a given project, the use of PyRDM requires minimal interaction from end-users who are running computational simulations. It is therefore expected that PyRDM could be used by other institutions to help address one of the psychological factors currently limiting the amount of software and data that is published along with journal articles, namely the lack of intrinsic motivation due to perceived lack of value from publishing software and associated data.

The application of the PyRDM library imposes some (albeit relatively minimal) implementation costs to produce the software-specific PyRDM-based publishing tools that are targetted towards end-users (i.e. the computational scientists actually running simulations). Therefore, encouraging and collaborating with other research software developers to integrate PyRDM into their research group's scientific software workflow will be an important step in the sustainability of PyRDM.

The interface to PyRDM has been designed to be as general and as minimalistic as possible. In the case of publishing software, only the software's name, version, category, and local Git repository location need to be provided. In the case of publishing data, only the file paths to the input/output data need to be provided along with a few basic publication details (e.g. title, description, and any tags). The Figshare authentication details are provided in the PyRDM configuration file which only needs to be set-up once, and all other parameters are determined automatically. These minimal requirements make PyRDM accessible to other scientific software stored in a GitHub repository.

Due to the vast number of ways that simulation codes and scientific software in general operate and manage/produce data, it was not possible to create a *generic* PyRDM-based publishing tool. For example, not all software uses an XML-based configuration file for simulation setups, so the paths to the raw data files may have to be obtained through some other method. Also, in the case of Fluidity, a file called "version.h" stores the version of the source code that was used to build Fluidity at compile-time. The version in this file is used in preference to the current version of the software's Git repository, since the Fluidity binary used to run a simulation may have been built from an older version. However, this method of determining the software's version may not be valid for another piece of software. It is therefore left to the developer to handle these issues on a software-specific basis. Nevertheless, PyRDM introduces a considerable amount of automation into the publishing process and into the workflow of scientific software in general.

In the future we aim to extend PyRDM to support other Git-based repository services such as Bitbucket (www.bitbucket.org), as well as Bazaar-based services such as Launchpad (www.launchpad.net). Should GitHub provide an API for creating a DOI for its repositories in the future, this functionality will also be incorporated into PyRDM. In addition, an interface to the Zenodo (www.zenodo.org) API is currently being developed in PyRDM to give users a choice of hosts/services with which to publish their data. Engagement with university libraries is also an important part of the long-term PyRDM roadmap, since many libraries run local repositories of publications produced by staff and students (e.g. the Spiral database at Imperial College London). Support for the popular DSpace framework (www.dspace.org) used to run these local repositories will therefore be introduced in a future version of PyRDM.

The field of computational science is in a relatively early stage with respect to dealing with the issues of recomputability and reproducibility. However, by combining ease-of-use with citable online repositories that can be referenced in future publications (thereby increasing the citability and recognition for the individual authors and research software developers), it is hoped that computational scientists will be further motivated to use automated publication tools such as PyRDM in the future.

**Acknowledgements**
This work was funded by Imperial College London. The authors would like to thank the two anonymous reviewers for their thoughtful comments.

**Funding statement**
This work was funded by Imperial College London.